\documentclass[aps,prl,twocolumn,groupedaddress,showpacs,showkeys]{revtex4-2}
\usepackage{graphicx}
\usepackage{dcolumn}
\usepackage{bm}

\begin{document}

\title{Magnetism of two-dimensional magnets in the presence of random fields}
\author{Essa M. Ibrahim$^1$}
\author{Ping Tang$^{1,2}$}
\author{Shufeng Zhang$^1$}
\affiliation{$^1$ Department of Physics, University of Arizona, 1117 E 4th Street, Tucson, AZ 85721}
\affiliation{$^2$ Beijing National Laboratory for Condensed Matter Physics, Institute of Physics, Chinese Academy of Sciences, Beijing, 100190, China }

\begin{abstract}
In low-dimensional magnets, thermal agitation and spatial disorders generate strong spin fluctuations that suppress the long-range magnetic ordering. We develop an analytical equation for the equilibrium magnetization of two-dimensional magnets at finite temperatures in the presence of the random magnetic field. We find that the random field induces a first-order phase transition in addition to the reduced Curie temperature. The first-order phase transition persists even in the presence of the moderate external magnetic field.
\end{abstract}

\date{\today}                     

\maketitle


In the last several years, research interests in low dimensional magnetism are getting greatly boosted by experimental identifications of many 2D magnetic materials, particularly, those of 2D van der Waals (vdW) magnetic semiconductors 
\cite{Huang,Gong,Ohara,Klein,Jiang, Lee, Huang2,Gong2}. Similar to the well-studied graphenes, the thickness of 2D vdW semiconductors can be precisely controlled down to one monolayer, and thus it is now possible that theoretical predictions on 2D magnetism can quantitatively compare with the experiments. Several magnetic and transport phenomena have already been reported experimentally, including the giant magnetoresistance \cite{Song}, spin-orbit torques \cite{Gupta, MacNeill, Alghamdi,Wang}, spin Seebeck, and Nernst effect \cite{Fang,Liu,Ito,Xu}. These experimental advances
raise an interesting perspective on 2D magnetic materials for spintronics applications. However, the qualitative explanation of
the experimental results is largely built on the theoretical models \cite{Zutic, Brataas, Bauer, current} that were developed for the 3D magnetic multilayers in the last two decades. It is unclear whether these 3D models are applicable for 2D magnets. 

On the other hand, theoretical research on magnetism and spin transport in low dimension (2D) has a long history, driven by fundamental physics in which many physical phenomena, e.g., critical exponents of phase transitions, depend qualitatively on dimensionalities \cite{Cortie}. One of most fundamental properties of low-dimensional magnetism is strong spin fluctuations. The thermal fluctuation at any finite temperature would destroy the long-range order for the isotropic Heisenberg models with short-range exchange interactions \cite{Mermin}. The fluctuation of random fields due to spatial disorders can also prevent the formation of the spatially uniform spontaneous magnetization in 2D, as shown by Imry and Ma \cite{Imry}. In comparison, 
the spin fluctuation in 3D does not in general destroy the ferromagnetism and the most of existing spintronic theories for spin
transport properties use the mean field approximation (MFA) to describe the equilibrium magnetization for 3D magnets. Since MFA completely fails for the 2D magnets, it is crucial to develop
an analytical theory for modeling equilibrium magnetization of 2D magnets before addressing non-equilibrium 
spin transport phenomena [1-8].

In this paper, we establish an analytical theory of the temperature dependence of the magnetization in the presence of the
random field. We show that, by using the self-consistent random phase approximation, the analytical form of the magnetization is simple enough and yet it has properly included the 2D spin fluctuation. We find that the random field not only reduces the ordering temperature through the reduced anisotropic gap, it also makes the phase transition from the second order to the first order. Even with a moderate external magnetic field, the first order transition persists.  

We start with the generic anisotropic Heisenberg Hamiltonian on a 2D square lattice,
\begin{equation}
    {\hat{\cal H}}=-J_{ex} \sum_{<i,j>}{\hat{\bf S}}_{i}.{\hat{\bf S}}_{j}- A \sum_{<i,j>} \hat{S}_{i}^z. \hat{S}_{j}^z-\sum_{i}(H + h_{i}) \hat{S}_{i}^z
\end{equation}
Where ${\hat{\bf S}}_{i}$ and $\hat{S}_{i}^z$ are respectively the spin and the $z$-component (taken as perpendicular to
the two-dimensional plane) of the spin operators at lattice site ${\bf R}_{i}$, $J_{ex}$ is the isotropic exchange integral, $A$ is the anisotropic exchange integral, $<ij> $ indicates the sum over nearest neighbors, and $ H$ and $ h_i $ are the external and the random magnetic field in the z-direction. The random field is assumed uncorrelated, i.e., the 
\begin{equation}
< h_i >_c = 0 ; \; \; \; \; < h_i  h_j >_c = \delta_{ij} \gamma^2 
\end{equation}
where $<>_c $ represents the configuration average over the distribution of the random field. The physical origins of the 
random fields may come from the spin-orbit coupling at the imperfect surface or interface in which the local electronic potential
is no longer periodic. For simplicity, we assume the strength of the random field, $\gamma$, is independent of the spin states and temperature. As the model Hamiltonian, Eq.~(1), has no exact solution even without the random field, one usually relies
on numerical methods such as quantum Monte Carlo simulation to determine the equilibrium magnetization and critical phenomena. As the analytical formulation for the magnetization is extremely useful for studying various spin transport properties, we will use a self-consistent random phase approximation (RPA) to determine the temperature dependence of the magnetization. Although the RPA is an approximate method, the physics of the spin fluctuation from the low energy excitations has been taken into account. 

To use the RPA, we first define the retarded Green's function of spin operators,
\begin{eqnarray}
    G^R_{ij}(t) =<<\hat{S}^+_i(t),\hat{S}^-_j>> \nonumber \\
\equiv i\Theta(t) <\hat{S}^+_i(t) \hat{S}^-_j>  -i\Theta(-t) <\hat{S}^+_i(t) \hat{S}^-_j>
\end{eqnarray}
Where $\hat{S}^{\pm} = \hat{S}_x \pm \hat{S}_y$ is lowering and raising spin operator, $\Theta(t)$ is the Heaviside step function and $<...>$ denotes the thermal average. The equation of motion for the above Green function is then
\begin{eqnarray}
i\frac{d G^R_{ij}(t)}{dt}=<[\hat{S}^+_i,\hat{S}^-_j]>\delta_{ij}+<<[\hat{S}^+_i(t),\hat{\cal H}],\hat{S}^-_j>> \nonumber
\end{eqnarray}
When we substitute Eq.~(1) into the commutator $[\hat{S}^+_i(t),\hat{\cal H}]$, the result contains the terms involving
the product of the three spin operators, e.g., $ \hat{S}^z_l\hat{S}^+_i \hat{S}^-_j $. To obtain a closed form for the 
Green's function, we use the RPA in which the longitudinal spin $\hat{S}^z_l $ and the transverse spin fluctuation $\hat{S}^+_i\hat{S}^-_j$ at the different sites $l\neq i,j$, are uncorrelated, i.e.,
\begin{equation}
<<\hat{S}^z_l\hat{S}^+_i ,\hat{S}^-_j >>=<\hat{S}^z_l><<\hat{S}^+_i,\hat{S}^-_j>> .
\end{equation}
Defining the site-independent magnetization $M(T) \equiv <\hat{S}^z_l>$, and making the Fourier transformation in space and time,
$G^R_{\bf kk'}(E) = (2\pi)^{-1} \sum_{ij} e^{i{\bf k} \cdot {\bf R}_i + i{\bf k}' \cdot {\bf R}_j }\int dt e^{-iEt} G^R_{ij}(t)$,
we find
\begin{eqnarray}
EG^R_{kk'}(E)=2M\delta_{kk'}+E^{(0)}_k  G^R_{kk'}(E)-\sum _q h_{k-q}G^R_{qk'}(E) \nonumber
\end{eqnarray}
where $E^{(0)}_k =  2zM[J(1-\gamma _k)+A] + H$ is the energy spectrum without the random
field, $z$ is the number of the nearest neighbors and $\gamma_k=\frac{1}{z}\sum e^{i\bf {k} \cdot \bf{R}}$ and 
the summation is over the nearest-neighbor sites. The above retarded Green's function can also be written in a compact form, 
\begin{eqnarray}
G^R_{kk'}(E)=\frac{2M\delta_{kk'}}{E-E^{(0)}_k-\Sigma(E,k)}
\end{eqnarray}
where $\Sigma(E,k)$ 
is the self-energy of the random field which can be expressed in terms of the series summation over the orders of the random field. If we keep the random
field up to the second order, the self-energy is
\begin{equation}
\Sigma (E,k)=\gamma ^2 \int \frac{g(\epsilon)d\epsilon}{E_k-\epsilon} 
\end{equation}
where we have used $<h_k h_{k'}>=\delta_{kk'} \gamma^2$ and $g(\epsilon)$ is the density of states. To further simplify the analytical
expression, we approximate the unperturbed dispersion by $E^{(0)}_k = M(8A +2Jk^2 a_0 ^2) + H $ for the square lattice such that the density of states is a constant for the energy within the magnon band, i.e., $g(\epsilon) = (8\pi J M)^{-1}$ for $ \Delta_0 < \epsilon < \Delta_0 + W_0$ where the energy gap is $\Delta_0 = 8MA +H$ and the bandwidth
$W_0 = 8 \pi JM$. The energy dispersion in the presence of the random field is given by the poles of Green's function, Eq.~(7). By explicitly
integrating the constant density of state in Eq.~(8), we obtain the energy dispersion with the random field,  
\begin{equation}
E_k= E^{(0)}_k+\frac{\gamma^2}{8\pi JM} \ln \left| \frac {E_k - \Delta_0}{\Delta_0 +W_0 -E_k} \right| 
\end{equation}
The above equation is an implicit equation that determines $E_k$ for a given magnetization $M$. However, $M$ is unknown
{\em a prior}, and must be determined self-consistently. Recall the spin operator identity, 
$ \hat{S}_i^z  =S(S+1)- (\hat{S}_i^z)^2 - \hat{S}^-_i \hat{S}^+_i$. For spin-1/2,  the identity becomes, 
$ \hat{S}_i^z =1/2- \hat{S}^-_i \hat{S}^+_i$ and thus, $M=1/2 - < \hat{S}^-_i \hat{S}^+_i >$. By taking the thermal averaging of the above identity, we have
\begin{eqnarray}
M=\frac{1}{2}-\sum_{kk'} \int \frac{dE}{2 \pi} \frac{2 {\rm Im} (G_{kk'}(E+i0^+))}{e^{\beta E}-1}
\end{eqnarray} 
By replacing  
\begin{eqnarray}
{\rm Im}G_{kk'}(E+i0^+)=2\pi M \delta \left( E-E_k^{(0)}-{\rm Re} \Sigma(E,k) \right) \nonumber\\ =2\pi M \delta(E-E_k)Z_k
\nonumber
\end{eqnarray}
where $Z_k=(1-\frac{\partial \Sigma}{\partial E_k})^{-1}$ into Eq.~(8), we have
\begin{eqnarray}
M=\frac{1}{2}-\int \frac{d^2k}{(2\pi)^2} \frac{2MZ_k}{e^{\beta E_k}-1}
\end{eqnarray}
Since we have used quadratic dispersion in the energy $E^{(0)}_k \propto k^2$,  we can change the integration over $d^2k$ to 
$dE_k$, i.e., replacing $d^2k=2\pi kdk=
\frac{\pi}{2MJ} dE^{(0)}_k =\frac{\pi}{2MJ} (1-\frac{\partial \Sigma}{\partial E_k})dE_k =\frac{\pi}{2MJ} Z_k^{-1} dE_k $ in Eq.~(9), we find
\begin{equation}
M=\frac{1}{2}-\frac{1}{4 \pi J} \left( \frac{1}{\beta} \ln \left| \frac{e^{\beta (\Delta +W_0)}-1}{e^{\beta \Delta}-1} \right| -W_0 \right)
\end{equation}
where the effective energy gap 
$ \Delta = \Delta_0 + (E_k-E^{(0)}_k) |_{k=0} $ and we have set the bandwidth $W_0$ unchanged since we assume the density of states remains unperturbed by the disorder.
By using Eq.~(7), we may explicitly write the effective gap,
\begin{equation}
\Delta = \Delta_0 - \frac{\gamma^2}{8\pi JM} \ln \left| 1+\frac{8\pi JM}{\Delta_0-\Delta} \right| .
\end{equation}
Equations (10) and (11) are our main results. The role of the random field is the reduction of the anisotropic gap 
from $\Delta_0$ to
$\Delta$. The limiting values of the gap reduction can be readily obtained from Eq.~(11). At the temperature well below the Curie temperature, $\gamma \ll 8\pi M J$, the above gap reduction $\Delta_0 - \Delta$ is negligible. As temperature increases, $M$ 
decreases and thus $\Delta_0-\Delta$ increases.  When $M$ becomes very small such that
$8\pi JM \ll \gamma$, the gap reduction reaches its maximum value of $\gamma$.
In the absence of the magnetic field, $\Delta_0 = 8MA$ decreases with temperature while the gap correction from
the random field increases with the temperature. Thus, at a certain temperature, the effective gap becomes too small to support
long-range ordering since the spin fluctuations at finite temperature destabilizes the magnetization and long range order is destroyed. More quantitatively, we shall numerically solve 
Eq.~(10), along with Eq.~(11), to determine the temperature dependence of the magnetization. 
 
\begin{figure}
    \centering
    \includegraphics[width=8.6 cm]{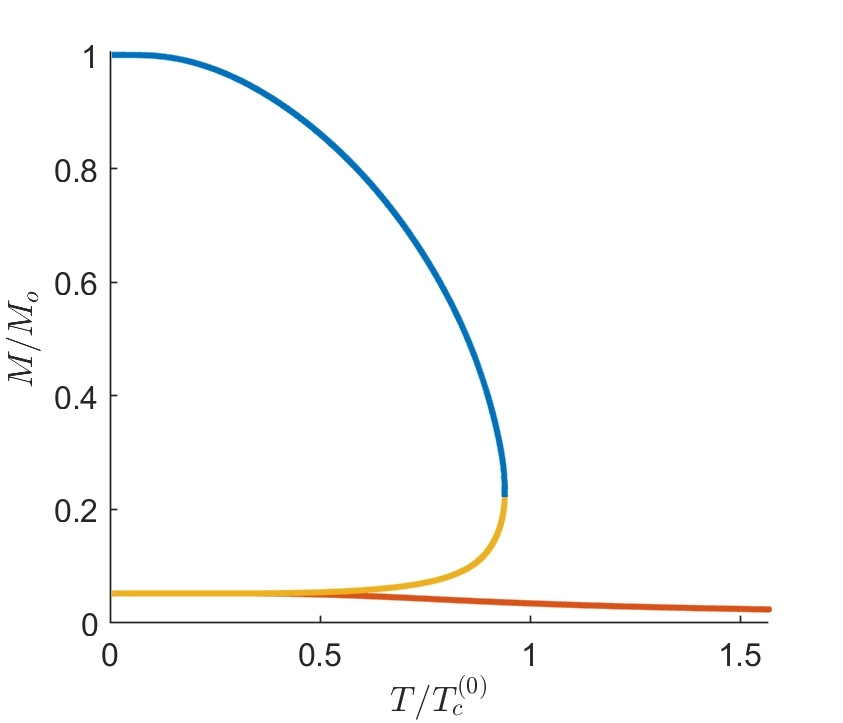}
    \caption{The complete solutions of $M$ as a function of temperature are calculated from Eq.~(10) and (11). The section 
with the blue line is the equilibrium ferromagnetic state while the other two sections are not stable states. The
section with the yellow color has higher free energy compared to the blue line and the section with the orange color has a negative effective energy gap. We have used $J=1$, $A=0.2$, $H=0$ and $\gamma =0.1 $}
    \label{fig:fig.1}
\end{figure}

We show general features of the mathematical solution of Eq.~(10) in Fig.1. For a given anisotropy constant $A$ and a random field strength $\gamma$, there are three solutions for the magnetization at the low temperature. The upper curve represents the
physically meaningful solution. The bottom curve is unphysical since it represents the case where the effective gap 
$\Delta$ becomes
negative. Clearly, the ground state is no longer in the $z$-direction when the gap is negative, and thus the magnon excitation
along the $z$-axis becomes invalid. The middle curve in Fig.~1 is also the solution of Eq.~(10), but the free energy is higher 
than the upper curve at the same temperature. Therefore, we will take the upper curve as the physical solution of the Eq.~(10) and we will only show the upper curve in the following numerical results.

\begin{figure}
    \centering
    \includegraphics[width=8.6 cm]{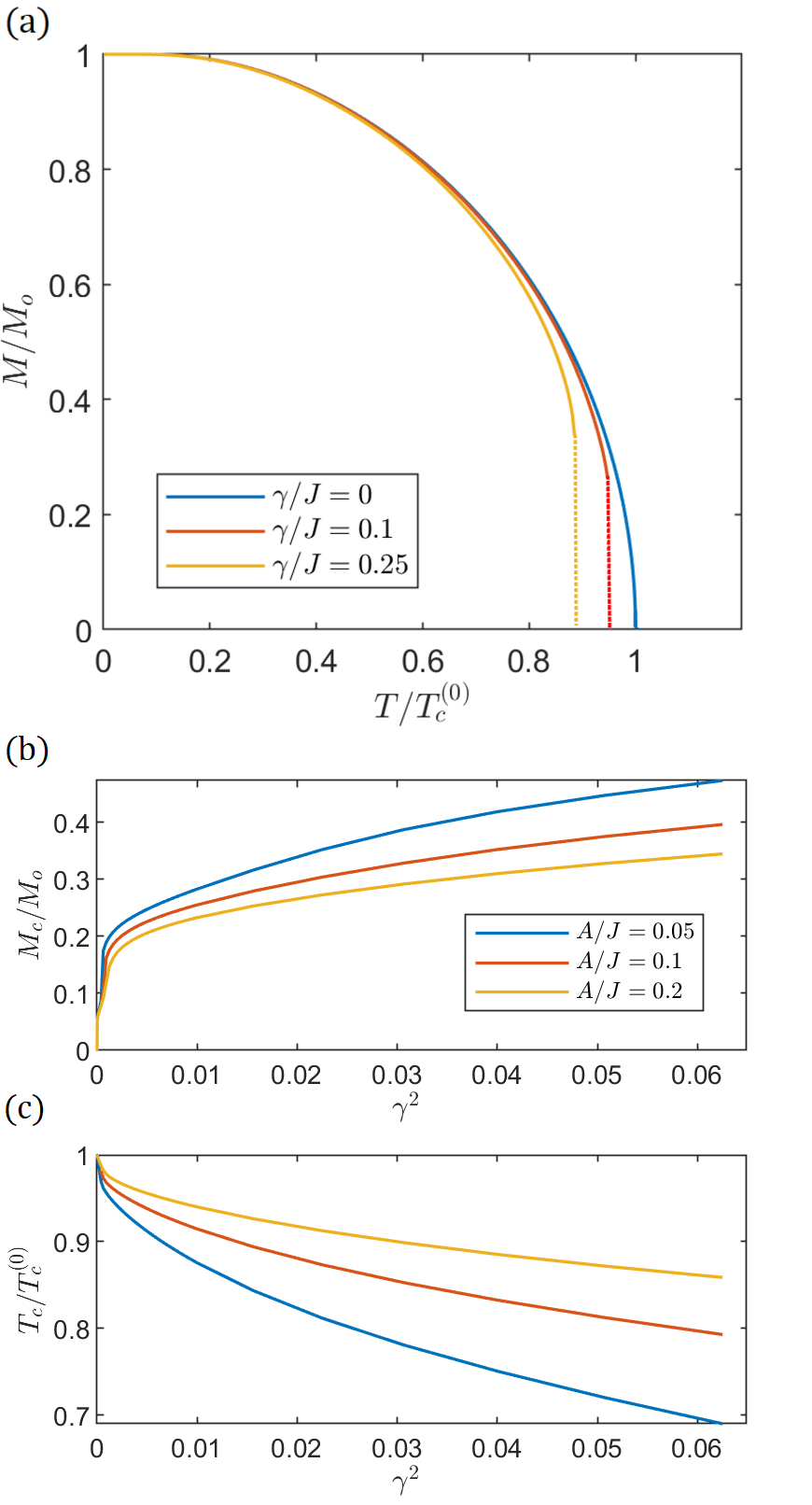}
    \caption{$(a)$ Magnetization as a function of temperature for several different values of the random fields without an external magnetic field. The dotted lines mark the critical points where the solutions end and the first-order phase transition occurs. We have used $J=1$ and $A=0.3$. $(b)$ and $(c)$ The critical magnetization and the critical temperature as functions of the strength of the random field for several systems with different anisotropy constants.}
    \label{fig:fig.2}
\end{figure}

We show the magnetization curves as a function of the strength of the disorder in Fig.~2. Without disorders, the magnetization
undergoes the second-order phase transition at the Curie temperature where the anisotropy gap can no longer stabilize 
the magnetization against thermal fluctuations. The
magnetization approaches zero at the critical temperature of the second-order phase transition, as shown in the blue line of Fig.~2. Two distinct features are seen as we increase the strength of the disorders. 
First, the reduction
of the transition temperature scales as the strength of the random field; this is expected since the effective gap, $\Delta$ of Eq.~(11), decreases as the random field increases. At low temperature, however, the effect of the random field is negligible since the gap $\Delta$ is not significantly different from $\Delta_0$. The second 
feature is more interesting: the phase transition becomes first order with the random field. When the temperature reaches a critical value, Eq.~(10) does not have a solution anymore,
indicating that the ferromagnetic phase we have assumed in deriving Eq.~(10) does not exist, i.e., the phase transition occurs at a finite value of the magnetization $M_c $ whose magnitude
scales with the strength of the random field, as shown in the insert of the Fig.2b. Since there are no solutions of Eq.~(10) for $T>T_c$, the magnetization is no longer uniform. Instead, the magnetization breaks into domains by the random 
field with the magnetization of each domain
fluctuating at high temperature, known as superparamagnetic (SPM) states. 

\begin{figure}
    \centering
    \includegraphics[width=8.6 cm]{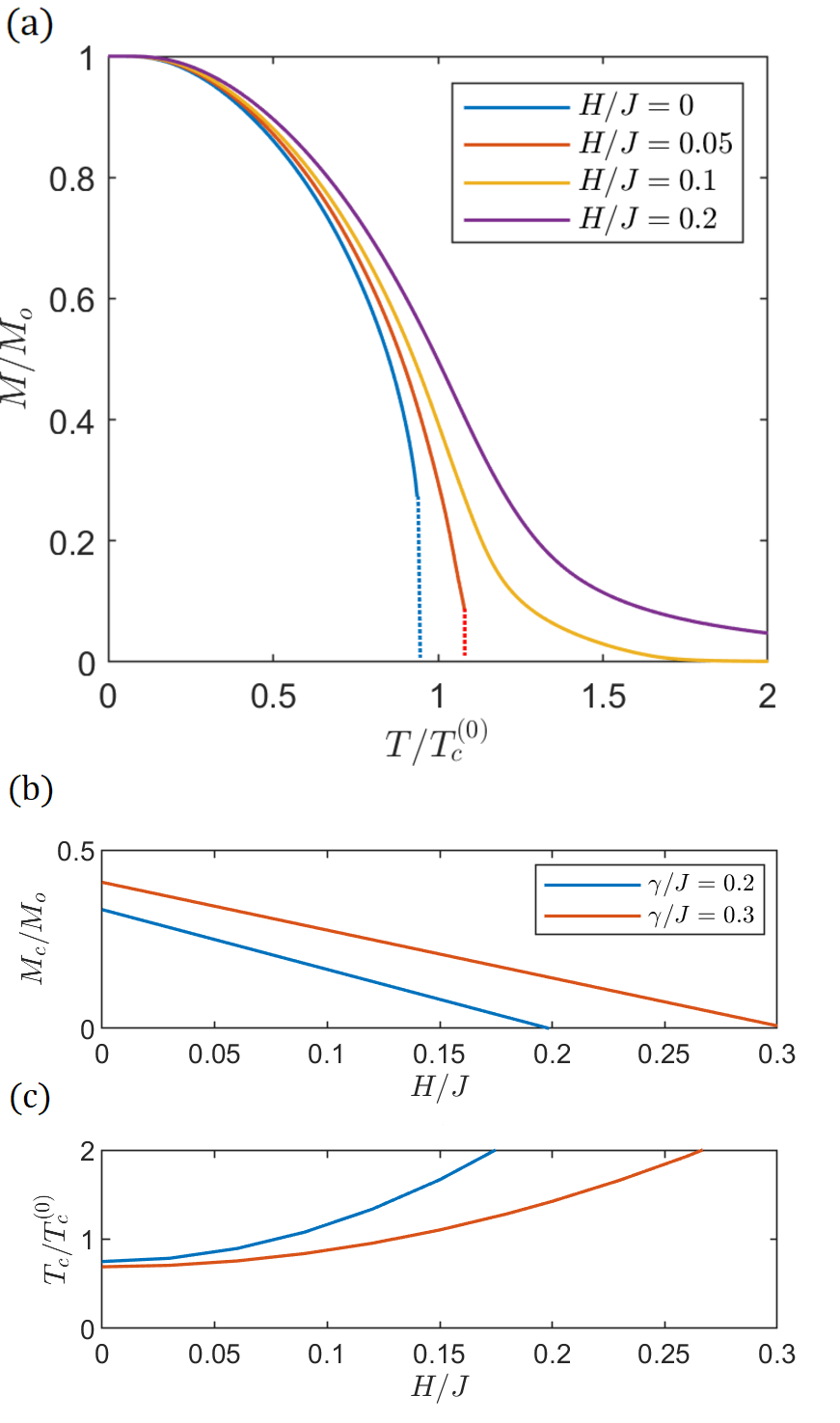}
    \caption{$(a)$ Magnetization as a function of temperature for several different external magnetic fields.The dotted lines mark the critical points where the solutions end and the first-order phase transition occurs. $A=0.2$ and $\gamma=0.1$. $(b)$ and $(c)$ The critical magnetization and the critical temperature as functions of the external magnetic field for several different random fields. $A=0.2$.}
    \label{fig:fig.3}
\end{figure}

\begin{figure}
    \centering
    \includegraphics[width=8.6 cm]{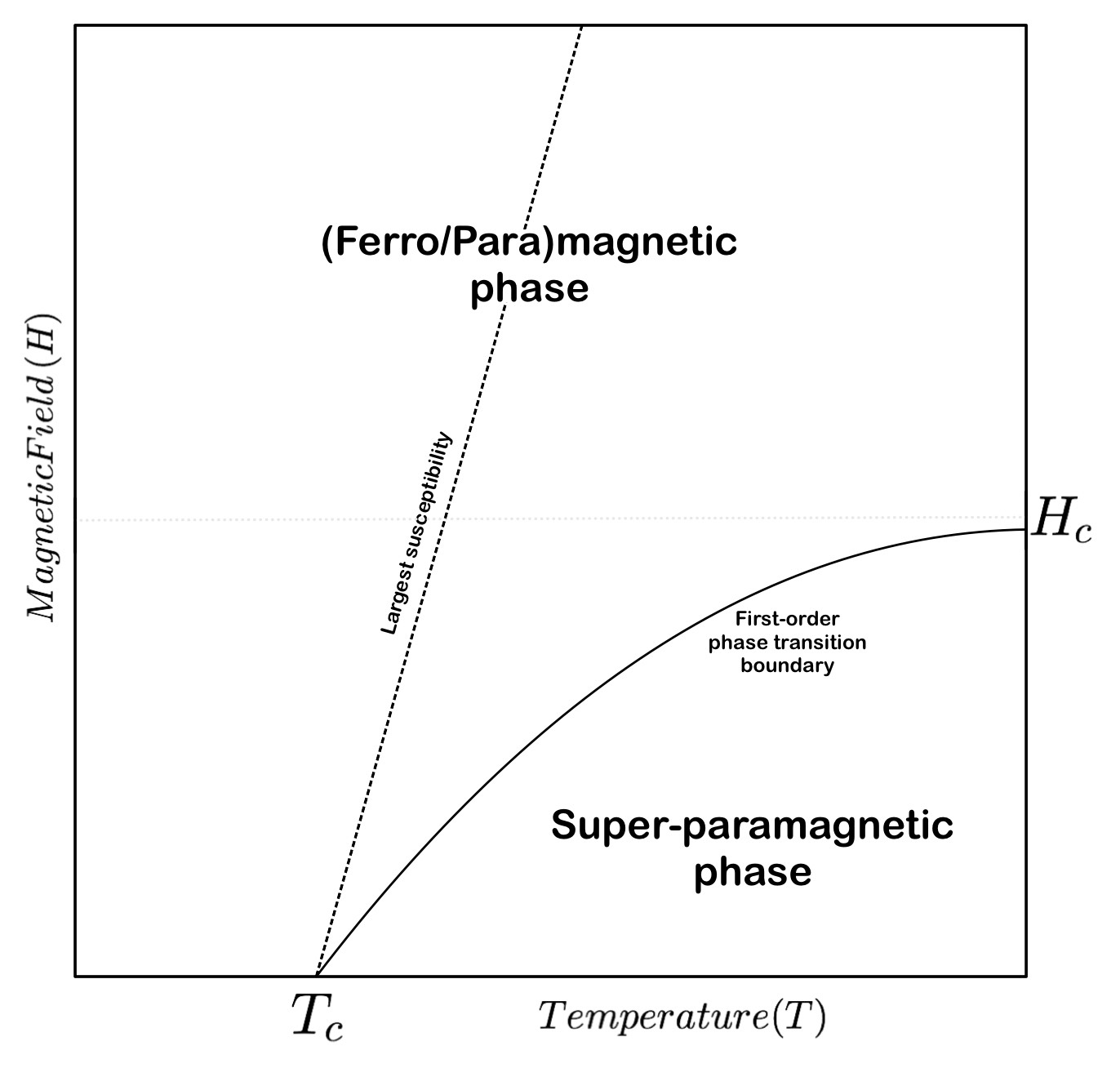}
    \caption{The phase diagram of a two-dimensional magnetic system with a random field.The solid black line represents the first-order phase transition from the (Ferro/Para)magnetic phase to the Super-paramagnetic (SPM) phase. The black dotted line represents the states of the largest susceptibility for a given temperature. One may define this dotted line as a ``boundary" 
between ferromagnetic and paramagnetic regions.}
    \label{fig:fig.4}
\end{figure}

We now discuss the effects of an external magnetic field on magnetization. Any external magnetic field breaks the time-reversal symmetry and thus the second-order phase transition which characterizes the transition between the time-reversal symmetry-broken and symmetry-conserving does not exist. With the random field, however, we find the first-order phase transition persists. If the magnetic field is
smaller or comparable to the strength of the random field, the solution of Eq.~(10) shows a similar first-order phase transition at a critical temperature. The explanation is as follows. The random field leads an effective
anisotropy gap as small as $\Delta = \Delta_0 -\gamma = 8 AM +H -\gamma$ when $M$ is small (or temperature is high). If $H $ is smaller than $\gamma$, the effective gap would be small or negative at the high temperature which leads to the collapse of the magnetization due to thermal fluctuation at a critical value of the temperature. In Fig. 3, we show the magnetization at several different external fields. When the external field is much larger than $\gamma$, the magnetization is essentially identical to that without the random field, i.e., there is no phase transition.

Taken together, we construct a temperature-magnetic field phase diagram in the presence of the random field, shown in 
Fig.~4. Since there is no second-order phase transition in the presence of the magnetic field, we denote the region 
as the (Ferro-Para)magnetic in which the 
solution of a uniform magnetization exists. When the magnetic field is small, there is a region in which there is no
solution with a uniform magnetization. While our theory cannot address the details of this non-uniform SPM, we postulate that the magnetization breaks into domains whose local directions 
are dictated by the net random field within the domains; this similar picture has been described by Imry and Ma \cite{Imry} 
from the viewpoint of thermal dynamics, and further confirmed by Fisher and co-workers \cite{Fisher}. When the magnetic field is large, the magnetic state is a single phase, however, we can artificially separate the regions with the ferromagnetic state FM
and the paramagnetic state PM by defining the boundary line as the maxima of the magnetic susceptibility.

This work was partially supported by the U.S. National Science Foundation under Grant No. ECCS-2011331.


\end{document}